\documentclass[conference]{IEEEtran}
\IEEEoverridecommandlockouts
\usepackage{times,eqnarray,amsmath,epsfig}
\usepackage{pgfplots}
\usepackage{algorithm}
\usepackage[noend]{algpseudocode}

\usepackage{booktabs} 
\usepackage{listings}
\usepackage{xcolor}
\usepackage[inline]{enumitem}
\usepackage{subfig}
\definecolor{mGreen}{rgb}{0,0.6,0}
\definecolor{mGray}{rgb}{0.5,0.5,0.5}
\definecolor{mPurple}{rgb}{0.58,0,0.82}

\begin{document}
\bstctlcite{IEEEexample:BSTcontrol}

%
\title{CIDPro: Custom Instructions for Dynamic Program Diversification}

\author{%
  \IEEEauthorblockN{%
  	Thinh Hung Pham\IEEEauthorrefmark{1},
    Alexander Fell\IEEEauthorrefmark{1},
    Arnab Kumar Biswas\IEEEauthorrefmark{2},
    Siew-Kei Lam\IEEEauthorrefmark{1}, and 
    Nandeesha Veeranna\IEEEauthorrefmark{1}
  }
  \IEEEauthorblockA{%
  	\IEEEauthorrefmark{1}Nanyang Technological University,  	Singapore\\
    Email: \{pham\_ht, afell, vnandeesha\}@ntu.edu.sg, siewkei\_lam@pmail.ntu.edu.sg
  }
  \IEEEauthorblockA{%
  	\IEEEauthorrefmark{2}University of South Brittany, France\\
    arnab-kumar.biswas@univ-ubs.fr
  }
}
 
\IEEEpubid{\begin{minipage}{\textwidth}\ \\[11pt]
		\_\_\_\_\_\_\_\_\_\_\_\_\_\_\_\_\_\_\_\_\_\_\_\_\_\_\_\_\_\_\_\_\_\_\_\_\_\_\_\_\_\_\_\_\_\_\_\_\_\_\_\_\_\_\_\_\_\_\_\_\_\_\_\_\\
		\textcopyright~Institute of Electrical and Electronics Engineers \\
		This is the author’s version of the research. It is for your personal use. \\
		The definitive version of the paper is in the proceeding of International \\
		Conference on Field Programmable Logic \& Application (FPL2018).
	\end{minipage}}

\maketitle

\begin{abstract}
	Timing side-channel attacks pose a major threat to embedded systems due to their ease of accessibility.
	We propose CIDPro, a framework that relies on dynamic program diversification to mitigate timing side-channel leakage.
	The proposed framework integrates the widely used LLVM compiler infrastructure and the increasingly popular RISC-V FPGA soft-processor. 
	The compiler automatically generates custom instructions in the security critical segments of the program, and the instructions execute on the RISC-V custom co-processor to produce diversified timing characteristics on each execution instance. 
	CIDPro has been implemented on the Zynq7000 XC7Z020 FPGA device to study the performance overhead and security trade-offs. 
	Experimental results show that our solution can achieve 80\% and 86\% timing side-channel capacity reduction for two benchmarks with an acceptable performance overhead compared to existing solutions. 
	In addition, the proposed method incurs only a negligible hardware area overhead of 1\% slices of the entire RISC-V system.
\end{abstract}


%
\IEEEpeerreviewmaketitle

\section{Introduction}
\label{Sec:Intro}
Embedded systems have become an integral part of our lives, and hence it is essential they are secure.
Unfortunately, the cryptographic schemes in embedded systems are deployed in unforeseen adversarial settings where keys can be compromised through side-channel attacks.
Such attacks have been reported on embedded devices \cite{Jude2015} where the attacker extracts secret information from a victim program by observing a physical phenomenon, e.g. execution time and power consumption, during its execution \cite{ge2016survey}.
In this paper, we consider the information leakage wherein the secret information of a victim program is unintentionally leaked to the attacker via the timing channel. 
For example, a modular exponentiation function (\emph{modExp}) used in RSA decryption (as shown in Algorithm~\ref{Alg:ModExp}), induces information leakage in execution time due to the absence of an else-branch, while the condition of the if-branch depends on the secret key $k$. 
The attacker can observe the execution time of the victim program and establish a correlation between the observation and hypothesis of the secret information such as a cryptographic key.
In this paper, we show that existing software countermeasures for mitigating timing leakage are ineffective especially when the cryptographic algorithms execute in bare metal mode or with an embedded OS (Operating System), which is common in embedded systems. 

To overcome the limitations of the existing software countermeasures, we propose a framework called \emph{CIDPro} that relies on dynamic hardware diversification to vary the timing characteristics of the program during execution. 
\emph{CIDPro} is built upon the widely used LLVM compiler infrastructure \cite{Pascal2015} and targets the Rocket core \cite{Krste2016} which is based on the RISC-V ISA. 
The framework is able to reduce timing side-channel leakage significantly by utilizing hardware diversification modules invoked by custom instructions that are inserted in the original source code.

The paper is organized as follows:
Section~\ref{sec:RelatedWork} discusses existing countermeasures for timing leakage reduction, while \emph{CIDPro} is presented in Section~\ref{sec:ProSol}.
In Section~\ref{sec:ResDis}, we provide a detailed discussion of the experimental results and Section~\ref{sec:conclude} concludes the paper.

\section{Related Work}
\label{sec:RelatedWork}
\begin{figure}[t]
	\centerline {\includegraphics [scale=0.5]{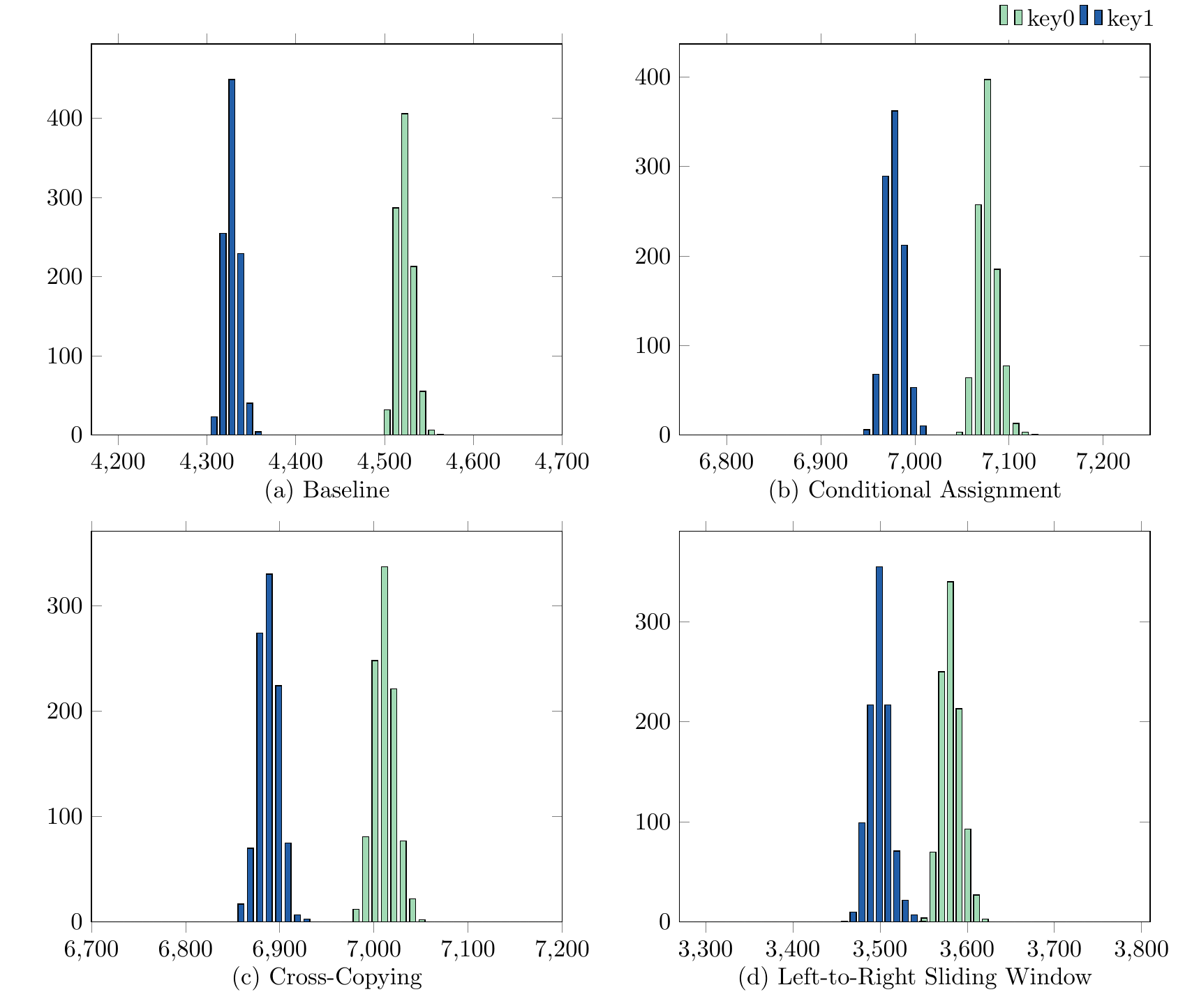} }
	\caption{Timing histogram of (a) a baseline program, (b) after transformation using cross-copying, (c) using conditional assignment and (d) using left-to-right sliding window.
		In each plot, the X-axis and Y-axis show the execution time in clock cycles and number of instances with that execution time, respectively.}
	\label{fig:timing_histo}
\end{figure}

A typical countermeasure against timing side-channel attacks is to apply program transformations such that the critical functions (e.g. \emph{modExp}) complete in constant time. 
Two representative program transformation techniques are \emph{cross-copying}~\cite{Molnar2006} and \emph{conditional assignment}~\cite{Agat2000}. 
In cross-copying, an else-branch is added with a dummy task that mimics the execution pattern in the if-branch to equalize the execution time of both branches.  
In conditional assignment, the condition is directly encoded into the branches using bit masks and bit-wise logical operators.

The countermeasures, mentioned above, suffer from several limitations: 
First, the program transformations are typically performed on high-level programs. As such, compiler optimizations may inadvertently reduce the effectiveness of the countermeasures, causing timing side-channel vulnerabilities at the micro-architecture level.
In addition, even though the conditional assignment removes critical conditions by flattening and converting them into primitive arithmetic and bitwise instructions, certain instructions (especially multiplications, divisions, etc.) may cause timing side-channel leakage due to variable clock cycle requirements based on the operand values.
Second, both the existing techniques cannot minimize the hamming weight of different keys.
Third, both the existing techniques result in performance degradation as additional instructions are introduced to equalize the execution times. 

\renewcommand{\algorithmicrequire}{\textbf{Input:}}
\renewcommand{\algorithmicensure}{\textbf{Output:}}
\begin{algorithm}
	\caption{Modular Exponentiation}
	\label{Alg:ModExp}
	\begin{algorithmic}[1]
		\Require{data $y$, private key $k$ of length $d$, integer $N$}
		\Ensure{$y^k \mod N$}
		\Procedure{modExp}{$y$, $k$, $N$}
		\State $r \gets 1$
		\For{$i \gets 1$ \textbf{to} $d$}
		\If {$(k\mod 2 == 1)$} 		\Comment{\emph{Critical Condition}}
		\State $r \gets (r\times y) \mod N$			
		\EndIf
		\State $y \gets (y\times y) \mod N$
		\State $k \gets k \gg 1$
		\EndFor
		\State \Return $r\mod N$
		\EndProcedure
	\end{algorithmic}
\end{algorithm}

The sliding window technique \cite{menezes1996} has been previously presented to reduce the effect of the hamming weight difference of keys on the execution time of the \emph{modExp} function. 
This technique not only improves the performance but also reduces the timing-channel leakage. 
However, the sliding window technique is not able to eliminate the information leakage.
In particular, the residual timing leakage is notable on embedded systems with a low noise runtime environment.

These limitations of the existing approaches are highlighted in Figures \ref{fig:timing_histo}(b)-(d), which show the timing histograms of the RSA algorithm (that includes the \emph{modExp} function) when it is executed with two different keys in the absence of an OS, i.e. in a bare metal environment that is common in embedded systems \cite{Michael2000}. 
The timing characteristics corresponding to the two different keys are clearly distinguishable with the existing approaches and hence, they do not provide effective countermeasures against timing side-channel attacks. 
As such, executing functions in constant time cannot be easily achieved in software, particularly for embedded systems.

Timing side-channel attacks to extract not only the Hamming weight but also the value of fixed Diffie-Hellman exponents and RSA keys, have been reported in \cite{Kocher1996}.
To mitigate the attacks, the authors presented blinding techniques to randomize the execution time.
However, the blinding technique requires a significant amount of computations, which poses a heavy burden for lightweight processors in embedded systems.

Hardware solutions usually require substantial changes to the processor architecture that impacts all programs running on the processor and also incur performance overhead on non-security critical programs. 
A secure processor that can protect against side-channel attacks using masking and hiding techniques, is proposed in \cite{Bruguier2016}.
Besides an independent data path to implement the masking scheme, a pipeline randomizer adds non-deterministic dummy control and data signals to the processor data path.
Another secure processor called Ascend, is presented in \cite{Ling2017} that relies on an ORAM controller to obfuscate the address bus.
However, information leakage cannot be prevented due to on-chip resource sharing. 
An instruction shuffler is proposed in \cite{Bayrak2012} to shuffle independent instructions randomly for protection against side-channel attacks.

In this paper, we propose the \emph{CIDPro} framework as a countermeasure against timing side-channel attacks especially in low noise embedded systems by utilizing both software and hardware methods with minimal resource overhead.

\section{Proposed Solution}
\label{sec:ProSol}

In this section, we describe the proposed \emph{CIDPro} framework and its hardware implementation that utilizes custom instructions to effectively reduce timing side-channel leakage with low hardware cost and acceptable performance overhead.

\subsection{CIDPro Framework}
\begin{figure}[t]
	\centerline{\includegraphics[scale=0.4]{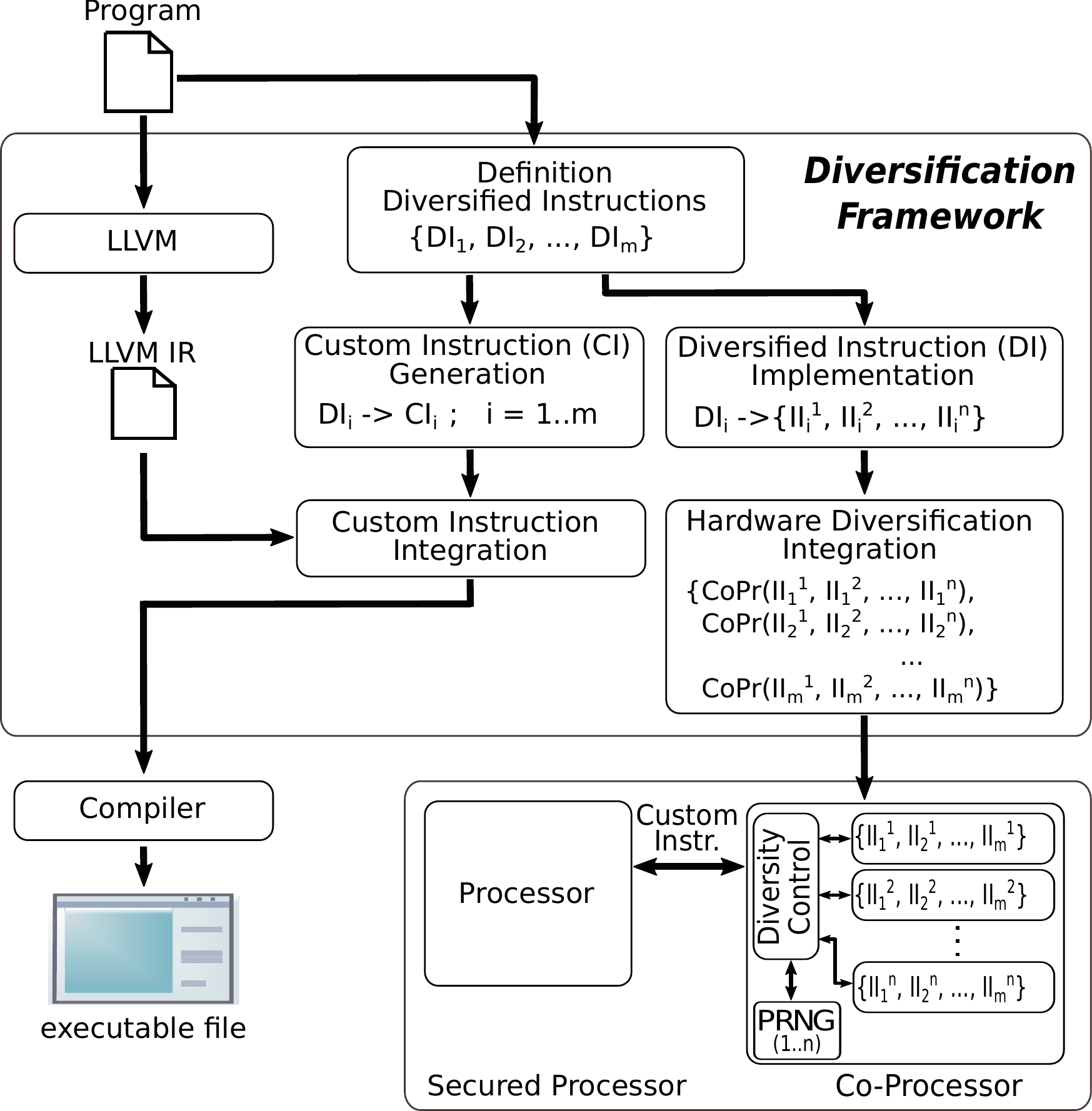}}
	\caption{Proposed \emph{CIDPro} Framework}
	\label{fig:ProFrm}
\end{figure}

The premise of our framework is hardware diversification using custom instructions to execute diversified (time-varying) instructions (DIs) as shown in Figure~\ref{fig:ProFrm}. 

The original source code of the program is given to the framework, which consists of two parts: 
\begin{enumerate*}\item An LLVM pass transforming the source code into an assembly file, and \item custom instruction generation of a set of $m$ diversified instructions $DI = \{DI_1, DI_2, \dots, DI_m\}$.\end{enumerate*}
Each $DI_i \in DI$ with $1 \le i \le m$ performs an arithmetic operation such as an ADD, MUL, etc, which is usually found in the source code of cryptographic primitives. 
Further, each $II_i^j \in DI_i$ represents a diversified version of the same operation with $1 \le j \le n$.
Therefore every instance $II_i^j$  provides the same functionality (i.e. $f(II_i^1) = f(II_i^2) = \dots = f(II_i^n)$), but exhibits different execution time characteristics.

For each $DI_i$, a corresponding hardware module $CoPr(DI_i)$ is implemented in a hardware description language (HDL).
These modules are integrated as custom instructions (CI) in the Secured Processor together with a Diversity Control Unit and a Pseudo Random Number Generator (PRNG).
At runtime, the Diversity Control Unit selects the corresponding $DI_i$ based on the CI, while the random number determines the $II_i^j$ to be executed.
Hence with every invocation of a CI, a different $II_i^j$ is selected, resulting in non-deterministic execution times of the program that reduces timing side-channel leakage.

The Custom Instruction Integration Unit facilitates the interfacing of LLVM and the set of DIs. An LLVM pass automatically identifies the arithmetic operations in the LLVM IR (Intermediate Representation) that exist in $DI$ and that are supported by the Co-Processor. The corresponding CIs are then integrated into the original source code.

\begin{figure}[t]
	\centerline{\includegraphics[scale=1]{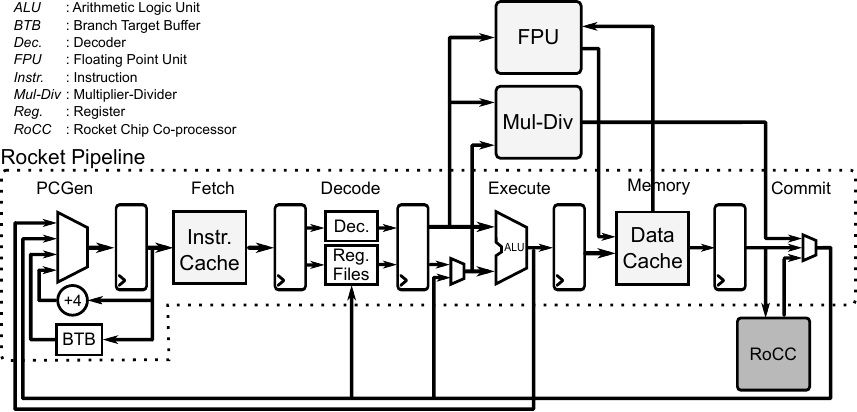}}
	\caption{Rocket Chip Micro-Architecture}
	\label{fig:microArch}
\end{figure}

Unlike existing hardware countermeasures, the proposed \emph{CIDPro} framework requires minimal changes to the processor architecture by utilizing a dedicated co-processor to execute CIs.
This leads to low resource overhead and design efforts.
In addition, the framework avoids negative side effects on non-security critical programs that run in the same environment as the security critical programs, since the former will be executed on the base processor unaffected.
Furthermore, a developer does not need to write programs in a new language or with security in mind.

The CIs are kept as private information and are automatically inserted into the security critical programs to mitigate the information leakage via timing side-channels. We assume that our proposed framework is not available to the
attacker and we also do not consider hardware probing or reverse engineering attacks through which the attacker can obtain details of the diversified hardware implementation. Finally, it is worth mentioning that while we have only considered single operations as custom instructions in this paper, the framework can be easily extended to generate time-varying custom instructions that consist of multiple operations.

\subsection{Hardware Implementation}
\label{subsec:Impl}

The Secured Processor shown in Figure~\ref{fig:ProFrm} is a RISC-V architecture that includes a Rocket core, a custom defined Rocket Chip Co-Processor (RoCC), L1 caches and a Floating Point Unit (FPU) \cite{Krste2016}.
The RoCC, which executes the CIs, is a tightly integrated extension to the processor pipeline as shown in Figure~\ref{fig:microArch}.
This enables the RoCC to stall the entire pipeline until the CI has completed its execution. In our work, the Secured Processor is implemented on the Zynq7000 XC7Z020 FPGA device.

\begin{figure}[t]
	\centerline{\includegraphics[scale=0.9]{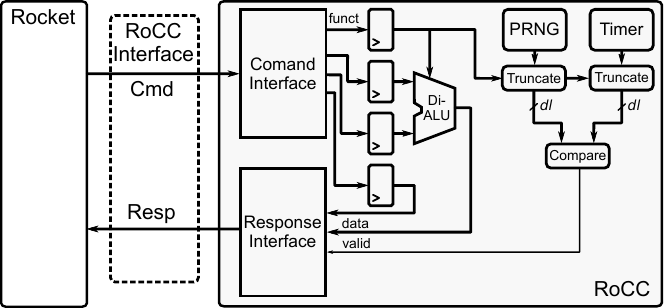}}
	\caption{Hardware Diversification on Co-processor}
	\label{fig:HardDiv}
\end{figure}

The Co-Processor utilizes the integrated DSPs on the FPGA fabric to support the operations required by the CIs.
Instead of implementing multiple versions of the same instruction for each $II_i^j \in DI_i$ with varying execution times, only one hardware module is implemented for $f(II_i^j)$ for all elements in $DI_i$. 
To emulate the different execution times of $II_i^j$, the PRNG is connected to a comparator that triggers a \emph{valid} signal when the random value matches the output of the timer. This indicates the completion of the operation on the Co-Processor and the valid result is passed to the Rocket core (refer to Figure~\ref{fig:HardDiv}).

The \emph{funct} signal determines the operation to be executed on the Diversifying ALU (Di-ALU) and also allows the developer to limit the range of numbers generated by the PRNG.
A higher range for the random value results in a larger reduction in the timing side-channel leakage. 
However, this also culminates in a longer overall execution time for the program, since the pipeline of the Rocket Chip stalls for long periods.
Hence a trade-off between channel leakage and execution time should be considered by setting an appropriate diversification level $dl$. 
The level specifies the number $n = 2^{dl}$ of diversified instructions of which the execution times are varied from 1 to $n$ clock cycles. $dl = 0$ means no diversification.
This trade-off is discussed in Section~\ref{subsec:Tradeoff}.

\section{Results and Discussion}
\label{sec:ResDis}

This section first describes the benchmark programs and the metrics used in our evaluations. This is followed by the investigation of the trade-off between performance overhead and information leakage reduction of the \emph{CIDPro} framework in a low-noise runtime environment (i.e. bare-metal mode) with respect to varying diversification levels.
Finally, the results of channel capacity reduction are presented to show the effectiveness of \emph{CIDPro} compared to existing solutions in both low-noise and OS environments (i.e. Linux mode).

\subsection{Benchmark Programs}
To compare the effectiveness of the proposed solution with existing methods, benchmark programs that can be implemented on embedded systems (like RISC-V based systems) that contain timing side-channel vulnerabilities are required. 
In our experiments, we use the RSA modular exponentiation (\emph{modExp}) to encrypt or decrypt a message \cite{Rivest1978} from the benchmark suite introduced in \cite{Heiko2015} and modular multiplication (\emph{mulMod16}) from the IDEA cipher \cite{Lai1992}.
Since these programs originate from practical cryptographic algorithms with different degrees of sophistication, they are meaningful candidates for the evaluations.

\subsection{Experimental Setup}

A timing side-channel is a communication channel created by unintentional information leakage by a victim program. 
Here the input is the set of values given to a victim program and the output is the timing observations of the program by an attacker. 
Like any other communication channel, timing side-channel can  be measured by Shannon's channel capacity \cite{Shannon1948} which represents the tight upper bound on the information transmission rate using that channel. 
We use a command-line tool called LeakiEst \cite{Tom2013} to estimate the side-channel capacity from observations of program execution times. 
Benchmark programs are executed on the Rocket chip system, and the execution time is extracted through the performance counters integrated in the Rocket chip. 
We collect 1000 samples for each of the two secret inputs for each benchmark program to obtain the leakage channel capacity using LeakiEst, which uses the iterative Blahut-Arimoto algorithm \cite{Blahut1972, Arimoto1972} to estimate the channel capacity.

We perform a distinguishing experiment by running each program with two distinct secret input values. 
Here the security concern is that the secret keys can be inferred from the execution time based on the bit patterns of the keys.
\begin{figure}
	\centering
	\includegraphics[width=0.957\columnwidth]{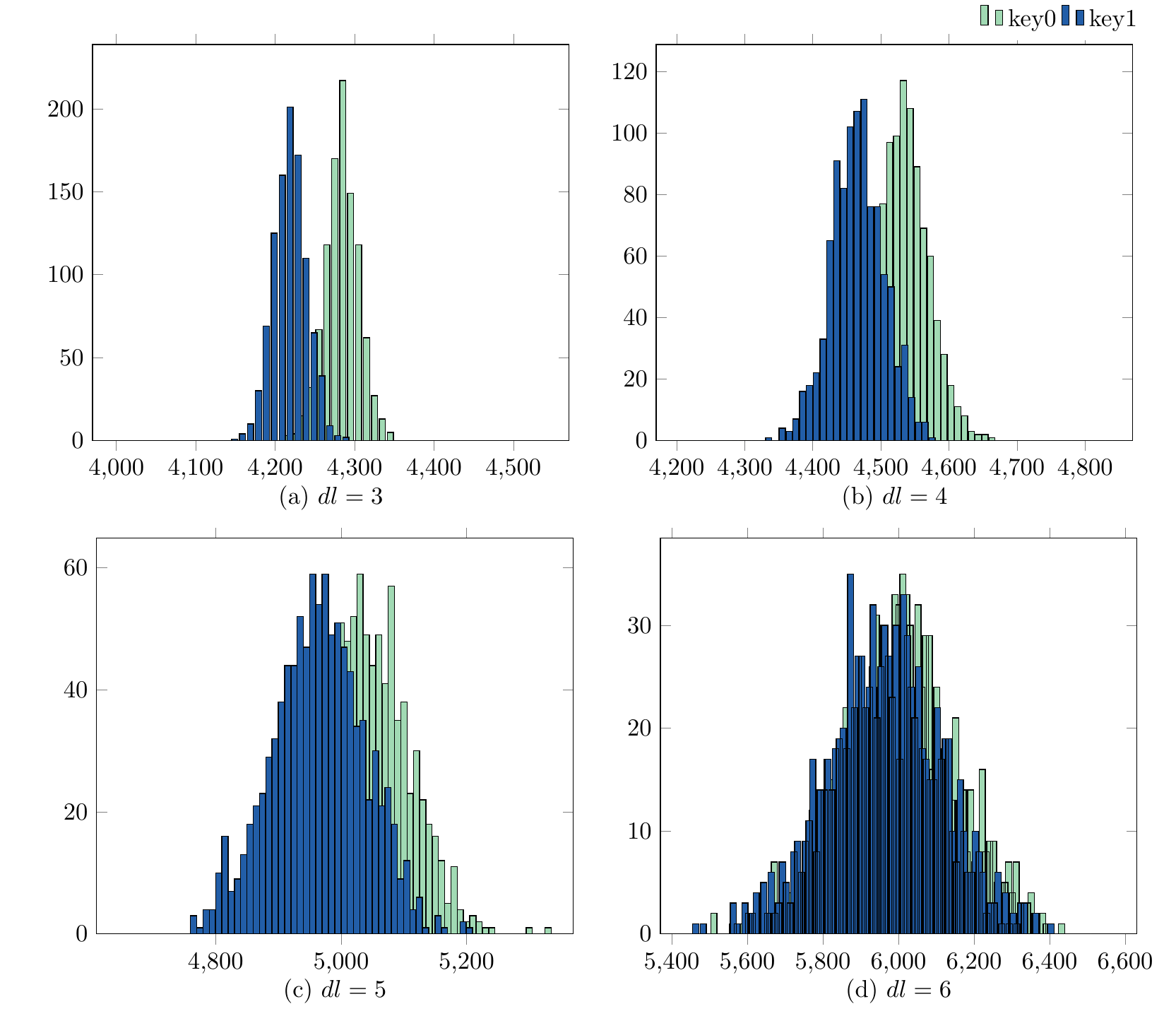}
	\caption{Timing histogram of \emph{CIDPro} for the \emph{modExp} benchmark with different diversification levels. In each plot, the X-axis and Y-axis show the execution time in clock cycles and number of instances with that execution time, respectively.}
	\label{fig:timing_histoPro}
\end{figure}
We compare our proposed solution with two existing and widely used solutions, i.e. cross-copying \cite{Agat2000} and conditional assignment \cite{Molnar2006}. 
In addition, we evaluated the sliding window technique for the modular exponentiation benchmark \cite{menezes1996}, which is widely used in cryptographic libraries \cite{bernstein2017} such as \emph{Libgcrypt}. 
Left-to-right (LR) windowed form with a window size of 3 is considered in our experiments.

\subsection{Performance overhead and security trade-offs}
\label{subsec:Tradeoff}

As mentioned in Section~\ref{subsec:Impl}, increasing the diversification level in \emph{CIDPro} increases randomness in the execution times of programs leading to a reduction in information leakage via timing side-channels.
Figure~\ref{fig:timing_histoPro} shows the timing histogram of the LR version of \emph{modExp} that utilizes the \emph{CIDPro} framework with different diversification levels.
As can be observed, if $dl$ increases, the distributions for two different keys (i.e. key0, key1) become indistinguishable which implies effective mitigation of information leakage via timing side-channels.

\begin{figure}[h]
	\centering
	\captionsetup[subfigure]{oneside,margin={2.35cm,0cm},singlelinecheck=false}
	\subfloat[Channel capacity\label{fig:ChCapvsRan_modExp}]{%
		\includegraphics[scale=0.62]{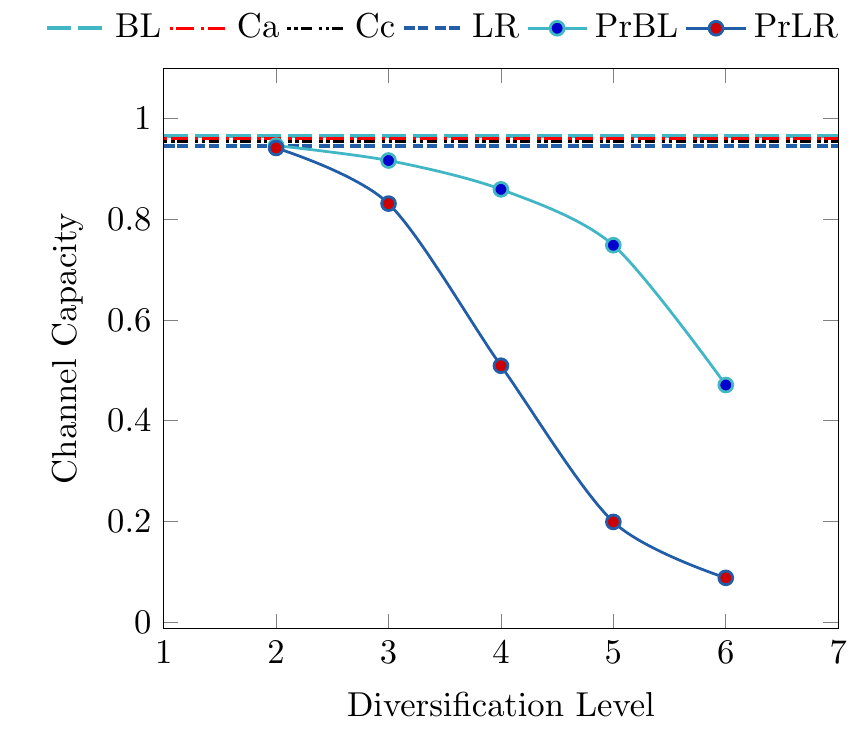}
	}
	
	\captionsetup[subfigure]{oneside,margin={2.05cm,0cm},singlelinecheck=false}
	\subfloat[Average execution time\label{fig:AvTimvsRan_modExp}]{%
		\includegraphics[scale=0.62]{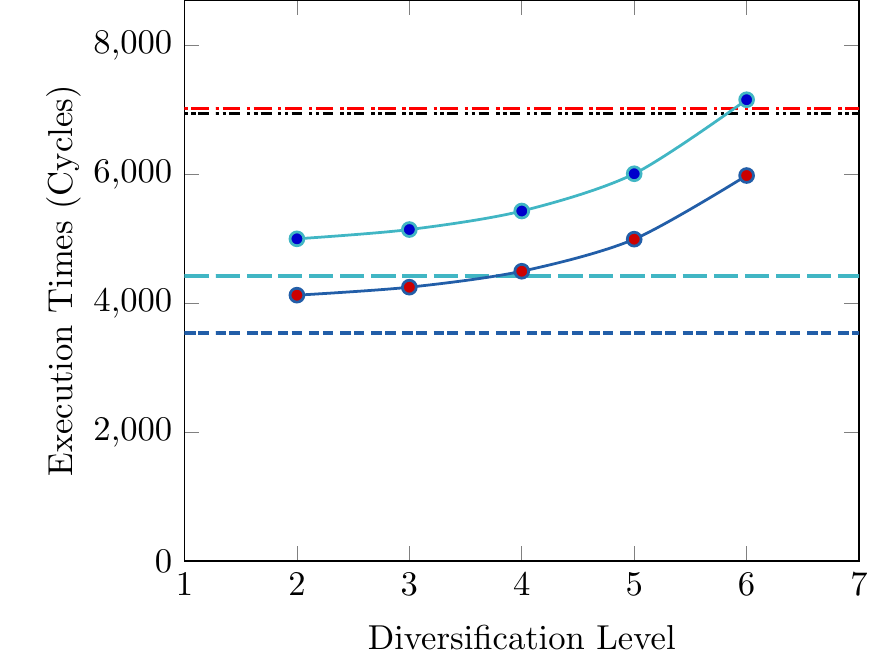}
	}
	\caption{Impact of various diversification levels on the \emph{modExp} benchmark in comparison to existing solutions}
	\label{fig:ChAvg_modExp}
\end{figure}

Figure~\ref{fig:ChAvg_modExp} shows the channel capacities and the average execution times of the \emph{modExp} benchmark using the \emph{CIDPro} framework with varying diversification levels compared to existing solutions. 
\emph{BL}, \emph{Ca}, \emph{Cc} and \emph{LR} denote the baseline program without countermeasures, and the program with the existing countermeasures i.e. conditional assignment, cross-copying, and left-to-right sliding window, respectively.
\emph{PrBL} and \emph{PrLR} represent the \emph{BL} and \emph{LR} programs that are modified by \emph{CIDPro}.
\begin{figure}[h]
	\centering
	\captionsetup[subfigure]{oneside,margin={2.3cm,0cm},singlelinecheck=false}
	\subfloat[Channel capacity\label{fig:ChCapvsRan_mulMod}]{%
		\includegraphics[scale=0.62]{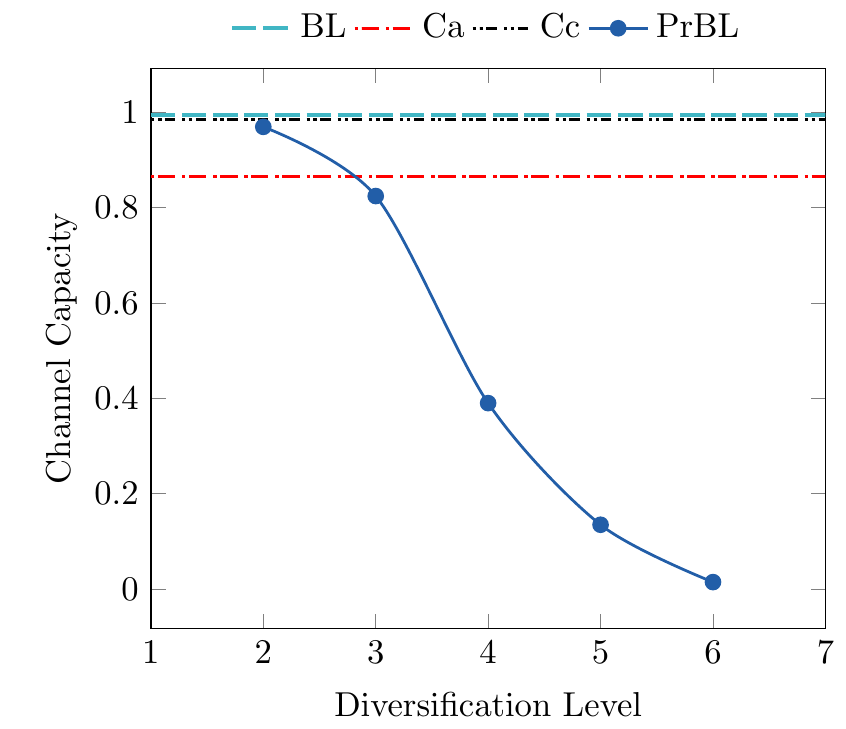}
	}
	
	\captionsetup[subfigure]{oneside,margin={2.05cm,0cm},singlelinecheck=false}
	\subfloat[Average execution time\label{fig:AvTimvsRan_mulMod}]{%
		\includegraphics[scale=0.62]{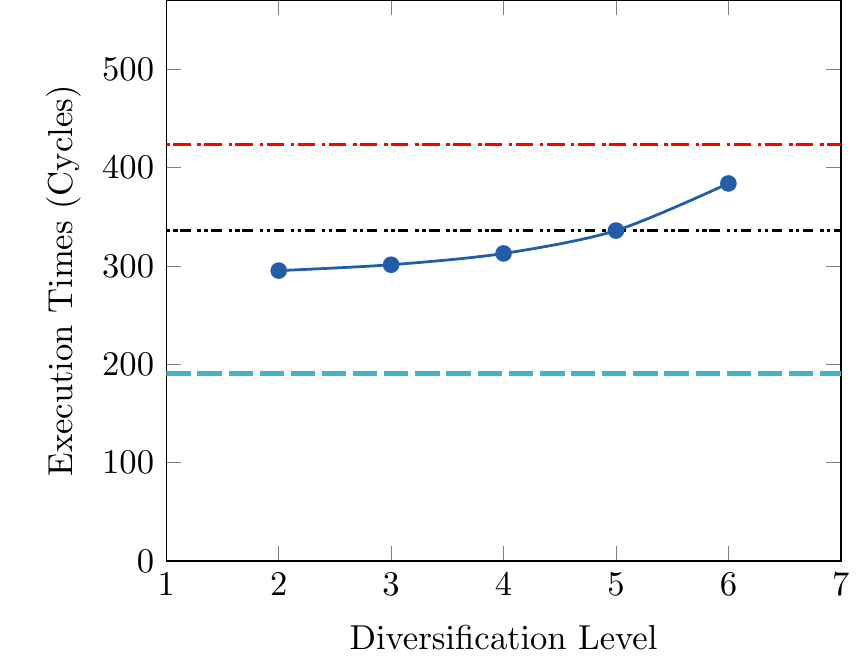}
	}
	\caption{Impact of various diversification levels on the \emph{mulMod16} benchmark in comparison to existing solutions}
	\label{fig:ChAvg_mulMod}
\end{figure} 
As can be observed in Figure~\ref{fig:ChAvg_modExp}(a), the channel capacities of \emph{BL}, \emph{Ca}, \emph{Cc} and \emph{LR} are almost equal to one, which indicates a high timing information leakage.
It is evident that the timing leakage reduces significantly for \emph{PrBL} and \emph{PrLR} with increasing diversification levels. In particular, \emph{PrLR} exhibits a higher reduction in channel capacity compared to \emph{PrBL} similar to the capacity comparison between \emph{LR} and \emph{BL} (refer to Figure~\ref{fig:timing_histo}).
Figure~\ref{fig:ChAvg_modExp}(b) shows that the execution times of \emph{Ca} and \emph{Cc} are notably higher compared to the baseline program, while \emph{LR} has the lowest execution time among all the techniques.
Although \emph{PrBL} and \emph{PrLR} result in an increased execution time compared to \emph{BL} and \emph{LR}, they are still faster than \emph{Ca} and \emph{Cc} for almost all the diversification levels considered. 
The increment in execution times of \emph{PrBL} and \emph{PrLR} in the range $2 \le dl \le 5$ is relatively small compared to those for $dl > 5$.

Figure~\ref{fig:ChAvg_mulMod} displays the result of the same experiment with the \emph{mulMod16} benchmark.
While \emph{Ca} results in a reduction in channel capacity, it has a significant increase in execution time compared to the baseline program.
\emph{Cc} exhibits a better performance than \emph{Ca}, but it is not able to reduce channel capacity in low-noise environments.
\emph{PrBL} has a significant reduction in channel capacity.
While the channel capacity reduces considerably with an increase in the diversification level, this adversely influences the execution time as expected.
For $dl \le 5$, the overall execution time for \emph{PrBL} remains lower than \emph{Cc} and \emph{Ca}.

\begin{figure}[h]
	\centering
	\captionsetup[subfigure]{oneside,margin={1.8cm,0cm},singlelinecheck=false}
	\subfloat[\emph{modExp}\label{fig:GainCost_mulMod}]{%
		\includegraphics[scale=0.505]{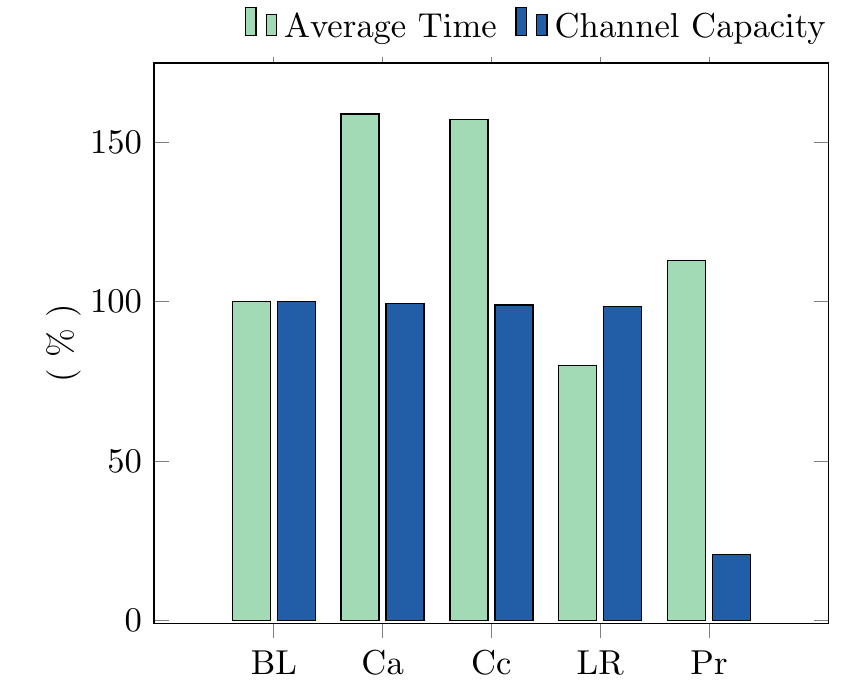}
	}%
	\captionsetup[subfigure]{oneside,margin={1.55cm,0cm},singlelinecheck=false}
	\subfloat[\emph{mulMod16}\label{fig:GainCost_mulMod}]{%
		\includegraphics[scale=0.505]{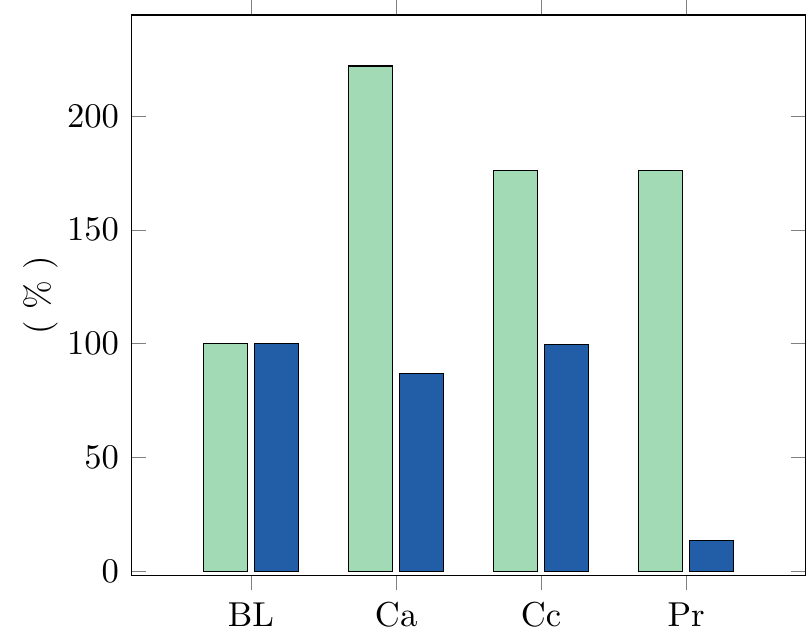}
	}
	\caption{Impact on the average execution time and channel capacity of \emph{CIDPro} in comparison to existing solutions and the baseline.}
	\label{fig:GainCost}
\end{figure}

\subsection{Evaluation Results}

Figure~\ref{fig:GainCost} compares the increased performance cost and the leakage reduction of the various countermeasures with respect to the baseline for the \emph{modExp} and \emph{mulMod16} benchmarks in bare-metal mode. 
The performance cost and the leakage reduction are evaluated in terms of the average execution time and the channel capacity, respectively.
\emph{Pr} denotes the programs that utilize the proposed \emph{CIDPro} with $dl = 5$ on \emph{LR} for the \emph{modExp} and on \emph{BL} for the \emph{mulMod16} benchmark.

As can be observed in the figure, the existing solutions have a negligible reduction in channel capacity, while \emph{Pr} achieves a significant reduction.
The channel capacity of \emph{Pr} is reduced to 20\% and 14\% for the \emph{modExp} and \emph{mulMod16} benchmarks, respectively.
Existing solutions and \emph{Pr} result in additional average execution time with the exception of \emph{LR} in the \emph{modExp} benchmark. 
The execution time increases due to the insertion of dummy instructions to balance branches in security critical conditions for existing methods and the long variable execution times of CIs.
For the \emph{modExp} benchmark, \emph{Ca} and \emph{Cc} increase the execution time by over 50\% compared to \emph{BL}, while the proposed solution requires only 13\% additional execution time.

\emph{Ca} and \emph{Cc} in the \emph{mulMod16} benchmark result in a significant increase in execution time of 120\% and 75\% respectively compared to \emph{BL}. 
Even though the proposed \emph{CIDPro} results in a similar execution time for \emph{mulMod16} as \emph{Cc}, it achieves a significant reduction in channel capacity.
It is worth mentioning that the timing cost in our proposed solution just applies to the critical functions that use the CIs to mitigate timing side-channel leakage.
The execution time of the normal functions remains unaffected.

\begin{figure}[t]
	\centerline{\includegraphics [width=4.1cm, height=2.79cm]{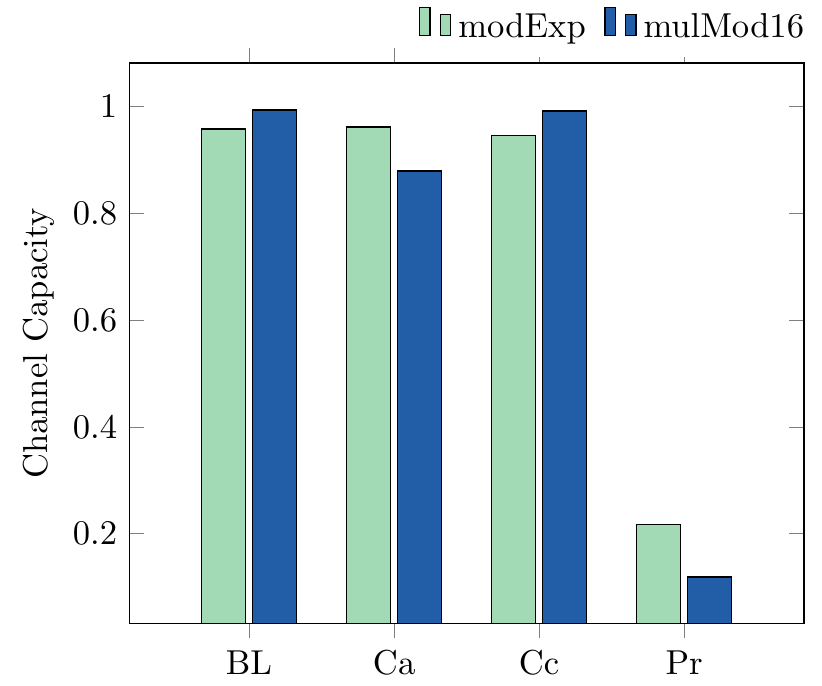}}
	\caption{Comparing channel capacity of the proposed solution and the existing solutions running on Linux OS environment.}
	\label{fig:ChCap_Lin}
\end{figure}

Figure~\ref{fig:ChCap_Lin} reports the results of the binaries executed in a Linux OS environment on the Rocket chip.
The Linux OS is very constrained with minimum effects on the programs and all executable files are compiled and linked statically. 
Therefore the timing noise level due to different auxiliary processes is expected to be lower in contrast to a fully functional OS. 
This means that the noise incurred by the Linux OS affects the information leakage only marginally.
As it can be observed in Figure~\ref{fig:ChCap_Lin}, the existing solutions (\emph{Ca} and \emph{Cc}) are not able to achieve notable reductions in channel capacity.
The results show that \emph{Pr} effectively mitigates the timing side-channel with channel capacity below 0.2 and 0.12 for the \emph{modExp} and \emph{mulMod16} benchmarks, respectively.

\subsection{Hardware Resource Utilization}
Table~\ref{tab:HarUtil} reports the hardware resource utilization of RoCC that implements the hardware diversification as part of \emph{CIDPro} compared to the typical Rocket chip system (\emph{RocketTile}) that includes a processor core (CPU) and a floating point unit (FPU). 
The hardware resource utilization is reported in terms of the number of slices and DSP blocks for FPGA implementation on the Zynq7000 FPGA device. 
The area overhead of RoCC is negligible (i.e. 1\% of slices) compared to the entire Rocket chip system.

\begin{table}[t]
	\centering
	\caption{Hardware Resource Utilization}
	\label{tab:HarUtil}
	\begin{tabular}{lcccc}
		\toprule
		\makebox[1.5cm]{}	&  \makebox[1.1cm]{RocketTile} & \makebox[1.1cm]{CPU} & \makebox[1.1cm]{FPU} & \makebox[1.1cm] {RoCC} \\
		\midrule
		\textit{Slices} 		& 6612 			& 1331	& 3458	& 84	\\
		\textit{DSPs}			& 34 			& 4		& 20	& 10	\\
		\bottomrule
	\end{tabular}
\end{table}

\section{Conclusion}
\label{sec:conclude}

We have proposed \emph{CIDPro}, a framework consisting of an LLVM compiler pass and hardware diversification to minimize timing side-channel leakage of cryptographic programs that run on soft-processors in embedded systems.
The proposed framework utilizes custom instructions to realize the hardware diversification without changing the base processor architecture and incurring only 1\%  area overhead for the co-processor. 
Experimental results targeting low noise runtime environments on RISC-V based system demonstrated that our framework achieves 80\% and 86\% timing side-channel capacity reduction compared to existing solutions. In addition, we show that the proposed solution leads to the best trade-off in timing leakage reduction and performance overhead compared to the existing countermeasures. 

\section*{Acknowledgment}
The research described in this paper has been supported by the National Research Foundation, Singapore under grant number NRF2016NCR-NCR001-006.



%

\bibliographystyle{IEEEtran}
\bibliography{IEEEabrv,fplref}

%






\end{document}